\documentclass[12pt,a4paper]{article}
\usepackage{latexsym,exscale,amssymb,bbm}
\usepackage{graphics,pstricks,pst-plot}
\usepackage[T1]{fontenc}
\usepackage{hyperref}
\usepackage[english]{babel}
\newcommand{\eqnn}[1]{\begin{eqnarray*}#1\end{eqnarray*}}

\newcommand{\eqnl}[2]{\begin{eqnarray}#1\label{#2}\end{eqnarray}}
\newcommand{\eqngrl}[3]{
\begin{eqnarray}#1\nonumber\\#2\label{#3}\end{eqnarray}}
\DeclareMathAlphabet{\boldmathe}{T1}{cmr}{bx}{it}
\newcommand{\mbfgr}[1]{\textit{\mbox{\boldmath$#1$}}}
\newcommand{\mbf}[1]{\boldmathe{#1}}
\newcommand{\refs}[1]{(\ref{#1})}
\newcommand{\ft}[2]{{\textstyle\frac{#1}{#2}}}
\font\af=msbm11
\def\mtxt#1{\quad\hbox{{#1}}\quad}
\def\pa{\partial}
\def\id{\mathbbm{1}}
\def\ha{\frac{1}{2}}
\def\tr{\hbox{tr}\,}
\def\dmur{d\mu_{\rm red}}
\def\al{\alpha}
\def\lam{\lambda}

\def\mbu{\mbf{j}}
\def\mbm{\mbf{m}}
\def\mbn{\mbf{n}}
\def\mbu{\mbf{u}}
\def\mb0{\mbfgr{0}}
\def\mbL{\mbf{L}}
\def\mbchi{\mbfgr{\chi}}
\def\mbvarphi{\mbfgr{\varphi}}

\def\N{\mbox{\af N}}

\def\Z{\mbox{\af Z}}
\def\cF{{\cal F}}
\newcommand{\Spf}{{\:\!\mbox{\scriptsize Spin}(5)}}
\hypersetup{%colorlinks,linkcolor=blue,citecolor=red,urlcolor=blue,
bookmarksopen,%
pdftitle={Ward Identities for Invariant Group Integrals},pdfauthor={S.~Uhlmann, R.~Meinel, A.~Wipf (FSU Jena)}}

\begin{document}
\begin{titlepage}
\begin{flushright}
hep-th/0611170
\end{flushright}
\par
\vskip .5 truecm
\centerline{\large\textbf{ 
Ward Identities for Invariant Group Integrals\footnote{Supported
by the Deutsche Forschungsgemeinschaft, DFG-Wi 777/8-2}}}
\par
\vskip 1 truecm
\begin{center}
\textbf{S.~Uhlmann, R.~Meinel and A.~Wipf}\footnote{e--mail:
s.uhlmann@uni-jena.de, meinel and wipf@tpi.uni-jena.de}\\[2mm]
\it{Theoretisch-Physikalisches Institut, \\
Friedrich-Schiller-Universit\"at Jena\\ 
Fr\"obelstieg 1, D-07743 Jena, Germany}\\
\vskip .5truecm
\end{center}
\par
\vskip 2 truecm

\begin{abstract}

\noindent
We derive two types of Ward identities for the generating functions
for invariant integrals of monomials of the fundamental characters for
arbitrary simple compact Lie groups. The results are applied to the
groups $SU(3)$, Spin$(5)$ and $G_2$ of rank $2$ as well as $SU(4)$.
\end{abstract}

{\small
}

\end{titlepage}
\setlength{\parindent}{0cm}

\section{Introduction}\label{intro}
Invariant group integrals are encountered in many problems in physics. Examples
are the one-link integrals in mean field or strong coupling expansions in Euclidian
lattice gauge theories \cite{creutz1} or mass-gap calculations in the Hamiltonian
formulation of these theories \cite{hamlattice}. They are used in the exact solution
of two-dimensional lattice gauge theories \cite{2dgauge}, the matching of gauge
theories to chiral models \cite{leutwyler} and the loop formulation of quantum
gravity on spin network states \cite{loopgravity}; they appear in random matrix
theory \cite{random} and its widespread applications in nuclear physics~\cite{nuclear},
quantum chaos and transport in mesoscopic devices~\cite{meso} as well as in quantum
information theory~\cite{quinfthy}. Apart from that, they are tightly connected
to various enumerative problems in mathematics such as counting the number of
invariants in a given tensor product of group representations or the number of
Young tableaux of bounded height~\cite{enumerative}.

We consider invariant integrals over compact Lie groups with left-right invariant
Haar measure $d\mu_{\rm Haar}$. For example, integrals of the form
\eqnl{
Z_G(j,j^\dagger)=
\int d\mu_{\rm Haar}(g)\exp\left(\tr (j^\dagger g)+\tr(jg^ \dagger )\right)}{allg1}
with a \emph{matrix valued} source $j$ can be calculated for
the $U(N)$ groups \cite{brower}. The solution of the Schwinger-Dyson
equations corresponding to the left and right group actions
on the source yields the closed form expression in terms of Bessel functions,
\eqnl{
Z_{U(N)}(j,j^\dagger)=\left(2^{N(N-1)/2}\prod_{m=0}^{N-1}m!\right)
\frac{\det \left(z_b^{a-1}I_{a-1}(z_b)\right)} {\det \left((z_b^2)^{ a-1}\right)},}{allg3}
where $(z_b/2)^2$ are the left- and right-invariant eigenvalues of
$j\cdot j^\dagger$. For the $SU(N)$ groups there are further invariants
independent of $j^\dagger$, for example $\det j$, and no comparable simple
solution to the Schwinger-Dyson equations are known.

An alternative approach based on an explicit parametrization
of the group elements has been proposed in \cite{eriksson}.
Here one ends up with series representations of the form
\eqnl{
Z_G(j,j^\dagger)=\sum_{0\leq n_1,\dots,n_s\leq\infty} 
a_{n_1,\dots,n_s}\,
x_1^{n_1}\cdots x_s^{n_s}\,,}{allg5} 
where $x_1,\dots,x_s$ are the algebraically independent 
invariants, $x_k\left(gjg'\right)=x_k(j)$. This method is only applicable
to small groups with sufficiently simple parametriziations. For example,
for $SU(2)$ one obtains the series 
\eqnl{
Z_{SU(2)}=\sum_{n=0}^\infty \frac{1}{n!(n+1)!}\left(\tr (jj^\dagger)+
\det j+\det j^\dagger\right)^n.}{allg7}
Already for $SO(3)$, the result is a triple series expansion
in powers of the three independent invariants.
For $SU(3)$, a convenient parametrization of the group elements
leads to the series \cite{eriksson}
\eqnl{
Z_{SU(3)}(j,j^\dagger)=
\sum_{n_1,\dots,n_4}\frac{2}{(n_1+2n_2+3n_3+n_4+2)!\:\!(n_2+2n_3+n_4+1)!}\,
\prod_{p=1}^4\frac{x^{n_p}}{n_p!}}{allg9}
with four left-right invariants
\eqnl{
x_1=\tr (jj^\dagger),\quad
x_2=\ft12\left(x_1^2-\tr (jj^\dagger)^2\right),\quad
x_3=\det(jj^\dagger),\quad
x_4=\det j+\det j^\dagger.}{allg11}
Since $x_4$ is not invariant under $U(3)$ transformations one finds a
triple series for this group. The corresponding results for $U(2)$ and
$U(3)$ actually coincide with the expression in terms of Bessel functions, eq.~\refs{allg3}~\cite{eriksson}.

Numerous results for so-called $n$-vector integrals containing only $n$
columns of unitary matrices were derived in~\cite{nvector} and more recently
in~\cite{aublam} using an elegant method based solely on the unitary
constraint and left-right invariance of the Haar measure. These methods
were applied to orthogonal groups in~\cite{braun}.

In~\refs{allg1}, $Z_G$ is a generating function for invariant integrals
of arbitrary functions on the group. In many applications one only needs
integrals of \emph{class functions}. Such functions are constant on
conjugacy classes,
\eqnl{
F(ghg^{-1})=F(h),}{einf0}
such that we can consider them as functions of the maximal Abelian torus
in $G$. By the Peter-Weyl theorem, the group characters form an orthonormal
basis on the Hilbert space of square integrable class functions. Characters
can be computed with the help of Weyl's character formula; they are
polynomials of the `fundamental characters' $\chi_p$ belonging to the
fundamental representations with highest weights $\mu_{(p)}$, $p=1,\dots,r$.

One of the authors was involved in the mean field analysis of effective
models for pure gauge theories near their phase transition point
\cite{meanfield}. There one is confronted with calculating invariant
integrals of the type
\eqnl{
Z_G(\mbu) = \int\dmur(g)\,\exp\big(\sum_{p=1}^r u_p\chi_{p}(g)\big),
}{einf3}
where $\dmur$ is the reduced Haar measure on $G$. Such invariant integrals
together with their Ward identities also show up in inverse Monte-Carlo
simulations where one calculates the couplings in the Polyakov loop
dynamics of gauge theories \cite{wozar}. Other applications are, e.g., the
computation of glueball masses in Hamiltonian $SU(N)$ lattice gauge theories
\cite{hamlattice, carlsson} or studies of the strong coupling limit of these
theories~\cite{kluberg}. The function $Z_G(\mbu)$ is the generating function
for the \emph{moments} of all fundamental characters,
\eqnl{
\frac{\partial^{m_1+\dots+m_r}}{\partial u_1^{m_1}\cdots \partial u_{r}^{m_r}}
Z_G(\mbu)\big\vert_{\mbu=0}
=\int \dmur(g)\, \chi_1^{m_1}(g)\cdots \chi_r^{m_r}(g)
\equiv t_{m_1,\dots,m_r}}{einf4}
for $r$-tuples $\mbu=(u_1,\ldots,u_r)$.\footnote{Note that for $j=u_1\id$,
eq.~\refs{allg1} yields the generating function for the defining representation
of $U(N)$ and its complex conjugate; however, for $N\geq 3$ there is no direct
way to derive the generating function $Z_{U(N)}(\mbu)$ for all fundamental
characters from the generating function $Z_{U(N)}(j,j^\dagger)$.}

In the literature there seem to be no suitable Ward identities for $Z_G(\mbu)$ 
for arbitrary groups~$G$. Thus we decided to publish our findings since they
could be useful for colleagues confronted with similar invariant integrals.

The paper is organized as follows: The next section gives an overview of the
known results for the generating functions $Z_G(\mbu)$ for $G=U(N)$ and
$G=SU(N)$ and extend those for $SU(N)$ in the special case of a generating
function for the defining representation. In section~\ref{genwardids} we derive
what we call geometric Ward identities, since they are based on the 
invariance of the Haar measure. These results hold for all compact and
simple Lie groups. In section~\ref{rk2} our method is applied to the
simple rank~2 groups $SU(3)$, Spin$(5)$ and $G_2$ and in section~\ref{rk3}
to $SU(4)$. In section~\ref{redHM} an alternative and more analytic method is 
presented which sheds further light on the properties of the reduced 
Haar measures. It turns out that the square of the Jacobian of the
transformation $\mbvarphi\mapsto\mbchi$ from the angles parametrizing the
maximal Abelian torus to the fundamental characters is proportional to
the density of the reduced Haar measure. We have no general proof of this
conjecture but have checked it for groups with rank $2$ and $3$. Based on
the conjecture we find alternative Ward identities which are applied to
the groups with rank $2$. In section~\ref{recrel} we use both types of
Ward identities to derive recursion relations for the moments
$t_{m_1,\dots,m_r}$ in \refs{einf4}. The appendices contain a detailed
description of the solution of the Ward identities derived in section~\ref{groupa2}
as well as tables of the lowest moments for the above rank~$2$ groups.
In the conclusions we comment on possible applications of our Ward
identities.

\section{Results for \texorpdfstring{$U(N)$}{U(N)} and \texorpdfstring{$SU(N)$}{SU(N)}}
Let $z$ be an arbitrary element of the \emph{group center}. By Schur's lemma
it acts on $\chi_p(g)$ by multiplication with a factor $z_p$ such that for the
$r$-vector $\mbchi=(\chi_1,\ldots,\chi_r)^t$,
\eqnl{
\mbchi'(g) = D(z)\mbchi(g)\mtxt{with}D(z)=
\hbox{diag}(z_1,\dots,z_r).}{cent1}
By invariance of the Haar measure under $g\mapsto zg$, this implies the symmetry
\eqnl{
Z_G(\mbu)=Z_G\left(D^{-1}(z)\mbu\right)\quad \forall z\in \hbox{center}.}{cent3}
This observation proves to be crucial for explicit computations of $Z_G$ in the following sections. We now briefly summarize the known results for $G=U(N)$ and extend those for
$SU(N)$ in the special case of a generating function for the defining representation.

For $U(N)$ the fact that the reduced Haar measure factorizes into a flat
measure times the absolute square of a Vandermonde determinant $\det\Delta(\varphi)$,
\eqnl{
\dmur(g) = \det\Delta(\varphi)\det\Delta^*(\varphi)\,d^r\varphi,}{einf5}
facilitates the integration such that the generating function
\eqnl{
Z_{U(N)}(u,v)=\int \dmur  \, \exp(u\,\tr g+v\,\tr g^*)}{einf7}
for integrals over the characters of the defining representation and its complex
conjugate can be computed explicitly. An integration over the maximal torus
parameterized by the angular variables $\varphi_1,\dots,\varphi_N$ yields the closed
expression
\eqnl{
Z_{U(N)}(u,v) = \det \pmatrix{
I_0 & I_1 & \cdots & I_{N-1}\cr
I_1 & I_0 & \cdots & I_{N-2}\cr
\vdots & \vdots & & \vdots\cr
I_{N-1} & I_{N-2} & \cdots & I_0}\left(2\sqrt{uv}\right).}{einf9}
In accordance with the general result \refs{cent3}, this function
is invariant under center transformations,
\eqnl{
Z_{U(N)}\left(e^{i\al}u,e^{-i\al}v\right)=Z_{U(N)}(u,v).}{cent5}
It can easily be checked that for $v=u^*$ this function is the limit of
\refs{allg3} for $z_b\to 4u^* u$; this limit has to be taken with caution 
since both the numerator and denominator in \refs{allg3} vanish for two
or more coinciding eigenvalues such that l'H\^opital's rule and certain Bessel
function identities are needed.

In order to calculate the analogous generating function for
$SU(N)$ we follow \cite{rossi} and insert the constraint $\det g=1$ on
the maximal Abelian torus in the form
\eqnn{
\delta_{\rm per}(\varphi_1 + \dots + \varphi_N)=\frac{1}{2\pi}\sum_n e^{in(\varphi_1 +
\dots + \varphi_N)}}
into the invariant integral \refs{einf7} and find
\eqnl{
Z_{SU(N)}(u,v)=
\sum_{n\in\Z}\left(\frac{u}{v}\right)^{Nn/2}\det \pmatrix{
I_n & I_{n+1} & \cdots & I_{n+N-1}\cr
I_{n-1}  & I_n & \cdots & I_{n+N-2}\cr
\vdots & \vdots & & \vdots\cr
I_{n-N+1} & I_{n-N+2} & \cdots & I_n}\left(2\sqrt{uv}\right),}{einf11}
see \cite{carlsson}. The generating function for $SU(N)$ is left invariant
by $\Z_N$ center transformations for which
\eqnl{
uv\to uv,\qquad \frac{u}{v}\to e^{4\pi i k/N}\, \frac{u}{v},\quad k=1,\dots,N.}{cent7}
In the case of $SU(2)$, the sum can be worked out and leads to
\eqnl{
Z_{SU(2)}(u,v)={_0F_1}\big[\,2\,\big|w^2\big] = I_0(4w)-I_2(4w),\quad w = \frac{u+v}{2}.}{einf13}
The case of $SU(3)$ is the subject of section~\ref{groupa2}. We are not aware
of similar explicit results for $SU(N)$ with $N\geq 4$. In these cases, the
complexity increases since the complex conjugate fundamental representations
are inequivalent.

In \refs{einf13}, the hypergeometric function ${_0F_1}$ does not appear accidentally;
for all $SU(N)$ the generating function for the defining representation,
\eqnl{
Z_{SU(N)}(u)=\int \dmur(g) \,\exp\left(u\,\tr g\right),}{einf17}
is one of the generalized hypergeometric functions
\eqnl{
{_pF_q}\Big[{a_1,\dots,a_p\atop b_1,\dots,b_q}\Big|x\Big]=
\sum_{n=0}^\infty \alpha_n\frac{x^n}{n!},\quad
\frac{\alpha_{n+1}}{\alpha_n}= \frac{(n+a_1)\cdots (n+a_p)}{(n+b_1)\cdots (n+b_q)},
}{einf15}
with $\alpha_0=1$. In the course of the paper, we will mostly identify them as
solutions to the generalized hypergeometric differential equation,
\eqnl{
\left\{\theta\prod_{i=1}^q (\theta + b_i -1)
-x\prod_{i=1}^p (\theta+a_i)\right\}{_pF_q}\Big[{a_1,\dots,a_p\atop b_1,\dots,b_q}\Big|x\Big]=0\mtxt{with}\theta=x\frac{d}{dx}.
}{einf16}
For $q=0$ (or $p=0$), the first (or second) products of differential operators are
to be replaced by the identity operator.

Center symmetry entails that the function $Z_{SU(N)}(u)$ in~\refs{einf17} is in
fact only a function of $x=\det(u\id)=u^N$, and $Z_{SU(N)}(u)=Z(x)$ satisfies the
differential equation
\eqnl{
\frac{d^N}{dx^N}\left(x^{N-1}Z(x)\right)=Z(x).}{einf21}
The solution is the hypergeometric function ${_0F_{N-1}}\big[\,2,3,\dots,N\big|x\big]$
such that
\eqnl{
Z_{SU(N)}(u)={_0F_{N-1}}\big[\,2,3,\dots,N\,\big|u^N\big].}{einf23}
Since for $SU(2)$, $\tr g=\tr g^\dagger$, this generalizes the standard result
\refs{einf13} (with $u=v$) by Arisue~\cite{arisue}. Eq.~\refs{einf23} follows
from \refs{einf11} when $v$ tends to zero so that
\eqnl{
Z_{SU(N)}(u)=\sum_{n\geq 0} u^{Nn} \det \Delta^{(n)},\quad
\left(\Delta^{(n)}\right)_{pq}=\left\{\begin{array}{c@{\hspace*{3mm}}l}
                                      \frac{1}{(n+q-p)!} & \mbox{for }n+q-p\geq 0,\\
                                      0 & \mbox{else}
                                    \end{array}\right. .
}{einf27}
If we multiply the $p$'th row of $\Delta^{(n)}$ with $(n+N-p)!$, it is
easy to calculate the determinants of these Toeplitz matrices,
\eqnl{
\det\Delta^{(n)}=\prod_{p=0}^{N-1} \frac{p!}{(n+p)!}\quad
\Longrightarrow\quad
\frac{\det\Delta^{(n+1)}}{\det\Delta^{(n)}}=\frac{1}
{(n+1)\cdots (n+N)}.}{einf29}
This proves eq.~\refs{einf23}.

%%%%%%%%%%%%%%%%%%%%%%%%%%%%%%
\section{Ward identities for generating functions}
\label{genwardids}
We denote the left derivative in the direction of the Lie algebra
element $T_a$ by $L_a$, i.e.\ $L_a f(g)=\frac{d}{dt}|_{t=0}f\big(\exp(it T_a)g\big)$
for some function $f$ on $G$. The Haar-measure is left (and right) invariant, thus
\eqnl{
\int d\mu_{\rm Haar}(g)\, (L_a f)(g)=0,\quad f\in L_2(G).}{ward1}
For class functions  $F$ and $\tilde{F}$ the function 
\eqnl{
\sum_a L_a\left(F\cdot L_a \tilde{F}\right)
\equiv \mbL\left(F\mbL \tilde{F}\right) = F\mbL^2 \tilde{F} + \mbL F\cdot \mbL
\tilde{F}}{ward3}
is a class function as well. In order to see this, we may assume that
$F$ and $\tilde F$ are basis elements (i.e., characters $\chi_\mu$ and $\chi_\nu$
of some representations with highest weights $\mu$, $\nu$). In $\sum_a (\chi_\mu
\cdot L_a^2\chi_\nu+L_a\chi_\mu\cdot L_a \chi_\nu)$, the first part of the sum is a
class function since the quadratic Casimir operator $\sum_a T_a^2$ commutes with
all group elements, and the second part is a class function since invariance
of the Killing metric $\tr T_a T_b$ under adjoint action by some group element~$h$
implies that $hT_a h^{-1}$ can be expanded as $R_a{}^c T_c$ with an orthogonal
matrix $R$.

Thus, \refs{ward1} with $f=F\cdot L_a\tilde{F}$ reduces to an integral over the
maximal Abelian torus,
\eqnl{
0 = \int \dmur\, \mbL\left(F\mbL \tilde{F}\right) =
\int \dmur\left(F\mbL^2 \tilde{F}+\mbL F\cdot \mbL \tilde{F}\right).}{ward5}
We take $\tilde{F}$ to be a fundamental character $\chi_p$ with
$p\in\{1,\dots,r\}$. The $\chi_1,\dots,\chi_r$ are good
coordinates for the maximal torus such that any class function 
can be thought of as a function of these characters,
$F=F(\chi_1,\dots,\chi_r)$. Then the identity \refs{ward5} reads
\eqnl{
0 = \int \dmur\left(F(\mbchi)\mbL^2\chi_p+\sum_q(\mbL \chi_p)
 \frac{\partial F(\mbchi)}{\partial\chi_q}(\mbL \chi_q)\right).}{ward7}
Every character $\chi_\mu$ of a representation $V_\mu$ with highest
weight $\mu$ is an eigenfunction of the Laplacian $\mbL^2$ with
eigenvalue $-c_\mu$, where $c_\mu$ is the value of the quadratic
Casimir in $V_\mu$,
\eqnl{
\mbL^2\chi_\mu=-c_\mu\chi_\mu.}{ward9}
To calculate the last term in \refs{ward7} we decompose the tensor product
of $V_\mu\otimes V_\nu$ into irreducible pieces,\footnote{The characters
$\chi_\lam$ are polynomials $\chi_\lam=\chi_\lam(\chi_1,\dots,\chi_r)$ of
the fundamental characters. For groups of higher rank it can be cumbersome
to calculate these polynomials.}
\eqnl{
V_\mu\otimes V_\nu=\bigoplus_{\lam} C^{\lam}_{\mu\nu}V_\lam,\mtxt{such that}
\chi_\mu\chi_\nu=\sum C^\lam_{\mu\nu}\chi_\lam.}{ward11}
Acting with $\mbL^2$ on this relation and using \refs{ward9}
we find the useful relation
\eqnl{
(\mbL\chi_\mu)\cdot (\mbL\chi_\nu)=\ha
(c_\mu+c_\nu)\chi_\mu\chi_\nu-\ha \sum_\lam C^\lam_{\mu\nu}\,c_\lam \chi_\lam.}{ward13}
with Clebsch-Gordan coefficients $C^\lam_{\mu\nu}$ and second
order Casimirs $c_\mu$. Now we may rewrite the Ward identity \refs{ward7}
as follows,
\eqngrl{
c_p\int\dmur\,\chi_p F(\mbchi) & = &
\ha\sum_{q=1}^r (c_p+c_q)\int\dmur\,\chi_p\chi_q\:\!
\frac{\pa F(\mbchi)}{\pa\chi_q}}{& & \!{} - \ha \sum_{q,\lam}
C^\lam_{pq}\,c_\lam \int d\mu_{\rm red}\,\chi_\lam\:\!
\frac{\pa F(\mbchi)}{\pa\chi_q},\quad 1\leq p\leq r.}{ward15}
We choose the class function $F=\exp(\mbu\cdot\mbchi)$ such that 
\eqnl{\int \dmur\, F = Z_G(\mbu),\quad
\frac{\partial F}{\partial \chi_q} = u_q F\mtxt{and}
\frac{\partial F}{\partial u_q}=\chi_q F.}{ward21}
Then, the identity \refs{ward15} translates into the following \emph{master equation}
for $Z_G$,
\eqngrl{
c_p\frac{\partial}{\partial u_p}Z_G(\mbu) & = &
\ha\sum_q (c_p+c_q) u_q\,\frac{\partial^2}{\partial u_q\partial u_p}Z_G(\mbu)}
{& &\!{}-\ha \sum_{q,\lam} C^\lam_{pq}c_\lam \, u_q\, \chi_\lam
\big(\mbfgr{\partial}\big) Z_G(\mbu),\quad p=1,\dots,r,
}{ward23}
where $\chi_\lam(\mbfgr{\partial})$ is the differential operator
obtained by formally evaluating the polynomial $\chi_\lam(\mbchi)$
at $(\pa/\pa u_1,\ldots,\pa/\pa u_r)$. The properties of the group $G$ enter
these Ward identities at three places: via the polynomials $\chi_\lam(\mbchi)$,
the values $c_\mu$ of the quadratic Casimir operators and the Clebsch-Gordan
coefficients $C^\lam_{pq}$. The coefficient functions of these linear partial
differential equations are constant or linear functions of the variables
$u_1,\dots,u_r$. Their complexity depends crucially on the polynomials $\chi_\lam$.
For $SU(3)$ and Spin$(5)$ all $\chi_\lam$ are quadratic, and one ends up with 
second order differential equations.

For explicit calculations, we have to fix our Lie algebra conventions which
were chosen to allow for an easy comparison with the computer algebra program
LiE~\cite{lie}. In these conventions, the Cartan matrix is given by
\eqnl{
K_{pq} = \frac{2(\al_{(p)},\al_{(q)})}{(\al_{(q)},\al_{(q)})}}{ward25}
in terms of the simple roots $\al_{(1)},\dots,\al_{(r)}$.
Thus, simple roots and fundamental weights are connected by the relation
\eqnl{
\al_{(p)}=\sum_q K_{pq}\mu_{(q)}.
}{ward27}
The \emph{shortest} simple root has squared length $2$. In particular, for
simply laced groups, the Cartan matrix reduces to $K_{pq}=(\al_{(p)},\al_{(q)})$.
Arbitrary roots and weights are linear combinations of the simple roots and
fundamental weights, respectively,
\eqnl{
\al=\sum m_p \al_{(p)}\equiv [m_1,\dots,m_r]\mtxt{and}
\mu=\sum n_p \mu_{(p)}\equiv [n_1,\dots,n_r].}{ward29}
The Weyl vector
\eqnl{
\rho=\ha \sum_{\al>0}\al=\sum_{p=1}^r \mu_{(p)}}{ward31}
plays an important role in the theory of representations; we will mostly need
it for its appearance in the formula for the value of the second order
Casimir operator in a representation~$V_\mu$,
\eqnl{
c_\mu=(\mu,\mu+2\rho).}{ward33}
With the help of \refs{ward27} and $(\al_{(p)},\mu_{(q)})=\delta_{pq}v_p$ 
the Casimir of the representation with highest weight $\mu=[n_1,\dots,n_r]$ 
can be written as
\eqnl{
c_\mu=\left(\mbn,K^{-1}\mbn'\right)\mtxt{with} n_p'=v_p(n_p+2).
}{ward35}
For simply laced groups all $v_p$ are equal to $1$ so that $n'_p=n_p+2$.
%%%%%%%%%%%%%%%%%%%%%
\section{Ward identities for groups of rank 2}\label{rk2}
In this section, we are going to investigate and exploit the
geometric Ward identities \refs{ward23} for the groups $SU(3)$,
Spin$(5)$ and $G_2$. For these rank~$2$ groups, the generating
function \refs{einf3} depends on two variables $u_1\equiv u$
and $u_2\equiv v$.
%%%%%%%
\subsection{The group \texorpdfstring{$SU(3)$}{SU(3)}}\label{groupa2}
As a simply-laced group $SU(3)$ has a symmetric Cartan matrix,
\eqnl{
K_{SU(3)}=\pmatrix{2&-1\cr -1&2},}{exa1}
and the quadratic Casimir of the representation with highest
weight $\mu=[n_1,n_2]$ is
\eqnl{
c_\mu=\frac{2}{3}\left(n_1^2+n_2^2+n_1n_2+3n_1+3n_2\right).}{exa3}
The fundamental 3-dimensional representation $3\equiv [1,0]$ and
its complex conjugate $\bar 3\equiv [0,1]$ both have Casimir $8/3$. Since
\eqnl{
3\otimes 3=6\oplus\bar 3,\quad
\bar 3\otimes\bar 3=\bar 6\oplus 3,\quad
3\otimes \bar 3=1\oplus 8,}{exa5}
the $\lam$-sum in \refs{ward23} contains both fundamental, two sextet and
the octet representations. The singlet representation has vanishing Casimir
invariant and does not contribute. The Casimir operator on the sextets
$6=[2,0]$ and $\bar 6=[0,2]$ takes the value $20/3$, and the octet $8=[1,1]$
has Casimir $6$. To derive explicit Ward identities we must express the
characters of the representations $6$, $\bar 6$ and $8$ in terms of the
fundamental characters. By~\refs{ward11}, eq.~\refs{exa5} yields
\eqnl{
\chi_6=\chi_3^2 - \chi_{\bar 3}^{\vphantom{2}},\quad
\chi_{\bar 6}=\chi_{\bar 3}^2 - \chi_3^{\vphantom{2}},\quad
\chi_8=\chi_3\chi_{\bar 3} - 1.}{exa7}
Thus the differential operators $\chi_\lambda(\mbchi)$ in the Ward identity \refs{ward23} 
read
\eqnl{
\chi_3 = \partial_u,\quad
\chi_{\bar 3} = \partial_v,\quad
\chi_6 = \partial^2_u - \partial_v^{\vphantom{2}},\quad
\chi_{\bar 6} = \partial^2_v - \partial_u^{\vphantom{2}},\quad
\chi_8 = \partial_u\partial_v - 1,}{exa9}
and the two Ward identities in \refs{ward23} for the generating function
\eqnl{
Z_{SU(3)}(u,v)=\int \dmur(g) \exp\big(u\chi_3(g)+v\chi_{\bar 3}(g)\big)
\quad\mbox{with}\quad \chi_3(g)=\tr g}{exa10}
have the simple form\footnote{The action of left-derivatives on characters
can also be worked out concretely in a matrix representation although this
requires more computational effort, cf.~\cite{WozarDiploma}.}
\begin{eqnarray}
\big(2u\partial^2_u + v\partial_u\partial_v + 8\partial_u - 6u \partial_v
  - 9v\big) Z_{SU(3)} & = & 0,\label{exa11}\\
\big(2v\partial^2_v + u\partial_v\partial_u + 8\partial_v - 6v \partial_u
  - 9u\big) Z_{SU(3)}&=&0.\label{exa13}
\end{eqnarray}
Note that by the property $d\mu(g)=d\mu(g^{-1})$ of the Haar 
measure, $Z_{SU(3)}(u,v)=Z_{SU(3)}(v,u)$ is a symmetric function. Thus it is
sufficient to study one of the two identities. In the defining representation
of $SU(3)$, the center $\Z_3$ acts by multiplication with $\exp(2\pi i/3)\id$,
and the symmetry \refs{cent3} reads
\eqnl{
Z_{SU(3)}\left(e^{2\pi i/3}u,e^{-2\pi i/3}v\right)=Z_{SU(3)}(u,v).}{exa15}
Together with the symmetry in its two arguments this suggests
that $Z_{SU(3)}$ is just a function of the combinations $u^3+v^3$ and $uv$.
In fact, we prove in the appendix that the solution to \refs{exa11} is given by
\eqnl{
Z_{SU(3)}(u,v)=\sum_{p,q=0}^\infty \frac{2}{(p+q+1)!\;\!(p+q+2)!\;\!q!}\,
{3(p+q+1)\choose p}(uv)^p \left(u^3+v^3\right)^q.}{exa19}
We have found only one comparably simple series representation
for the generating function in the literature \cite{carlsson}. To arrive at their
result the authors took the quartic series~\refs{allg9} and calculated $2$
of the $4$ infinite sums. It seems evident that our method based on geometric Ward
identities is more efficient to find simple series representation for generating
functions.

As two special cases, we will restrict $Z_{SU(3)}(u,v)$ to the $u$-axis (i.e., $v=0$) and
the diagonal (i.e., $u=v$). Instead of performing a resummation of the general result
\refs{exa19}, we derive differential equations for $Z_{SU(3)}$ in these cases. First
we solve the Ward identities (\ref{exa11},\ref{exa13}) on the $u$-axis, where
\begin{eqnarray}
\big(2u\partial^2_u + 8\partial_u Z - 6u\partial_v\big)Z_{SU(3)}\vert_{v=0}=0,\quad
\big(u\partial_v\partial_u + 8\partial_v - 9u\big)Z_{SU(3)}\vert_{v=0}=0.\label{exa21}
\end{eqnarray}
To get rid of the $v$-derivatives at $v=0$ we act with $u\partial_u$ on the
first equation and use \refs{exa21} to eliminate the terms $\partial_v Z_{SU(3)}$ and
$\partial_u\partial_v Z_{SU(3)}$. We find the ordinary differential equation
\eqnl{\left(u^2\partial^3_u+12 u\partial^2_u+28\partial_u-27u^2\right)
Z_{SU(3)}(u,0)=0.}{exa23}
Since $Z_{SU(3)}(u,0)$ only depends on $u^3$ we may equally well use $x=u^3$ as a
new variable, $Z_{SU(3)}(u,0)=Y_{SU(3)}(x)$. Then the differential equation takes the simpler form
\eqnl{\left(x^2\partial_x^3+6x\partial_x^2+6\partial_x-1\right)Y_{SU(3)} =
(x^2Y_{SU(3)})'''-Y_{SU(3)}=0,}{exa25}
which has a solution in terms of a generalized hypergeometric function (cf.\ 
\refs{einf21}), $Y_{SU(3)}(x)={_0F_{2}}\big[\,2,3\,\big|x\big]$.
This is just the result \refs{einf23} for the group $SU(3)$.

In order to find a differential equation on the diagonal $u=v$, we act by the operator
$(3u\pa_u + 5u\pa_v + 5 + 4u)$ on \refs{exa11} and evaluate the result at $u=v=:t$.
In terms of 
\begin{eqnarray}
  \frac{d}{dt}Z_{SU(3)}(t,t) &\!\!\!=&\!\!\!(\pa_u+\pa_v)Z_{SU(3)}, \qquad
  \frac{d^2}{dt^2}Z_{SU(3)}(t,t) =
    (2\pa_{uu}+2\pa_{vv})Z_{SU(3)},\nonumber \\
  \frac{d^3}{dt^3}Z_{SU(3)}(t,t) &\!\!\!=&\!\!\!(2\pa_{uuu}+6\pa_{uuv})Z_{SU(3)},\label{exa27}
\end{eqnarray}
the resulting ordinary differential equation reads
\eqnl{
\Big( t^2\frac{d^3}{dt^3} + t(10-t)\frac{d^2}{dt^2} - 2(12t^2+t-10)\frac{d}{dt}
- 12t(3t+5)\Big) Z_{SU(3)} = 0.}{exa29}
This is solved by a function
\eqnl{Z_{SU(3)}(t,t)=1+t^2+\sum_{n=3}^\infty \frac{a_n}{n!} t^n}{exa31}
where the coefficients $a_n$ satisfy the recursion relation
\eqnl{a_{n+1} = \frac{1}{(n+4)(n+5)} \big(n(n+1)a_n - 12n(2n+3)a_{n-1} -
36n(n-1)a_{n-2}\big) = 0.}{exa33}
Together with $a_0=1$, $a_1=0$, and $a_2=2$, this determines all coefficients in
the expansion~\refs{exa33}.

%%%%%%%%%%%%%%%%%%%%%%%%%
\subsection{The group \texorpdfstring{Spin$(5)$}{Spin(5)}}\label{groupb2}
Since Spin$(5)$ is not simply-laced, the Cartan matrix
\eqnl{
K_\Spf=\pmatrix{2&-2\cr -1&2}}{exb1}
is not symmetric. With the convention in \refs{ward25}, we take $\al_1$ to be the
longer root, $\al_1^2=4$ and $\al_2^2=2$. The eigenvalue of the Casimir operator
of the representation with highest weight $\mu=[n_1,n_2]$ reads
\eqnl{
c_\mu=2n_1^2+n_2^2+2n_1n_2+6n_1+4n_2.}{exb3}
The fundamental representations are the $SO(5)$ vector representation 
$5=[1,0]$ and the spin representation $4=[0,1]$. The center $\Z_2$ is
generated by $-\id$ in the spin representation and acts trivially in the
vector representation; the center symmetry~\refs{cent3} implies that
\eqnl{
Z_\Spf(u,v)=\int \dmur(g) \exp\big(u\chi_5(g) + v\chi_4(g)\big)
= Z_\Spf(u,-v)}{exb4}
is an even function in $v$. For the geometric Ward identities we need
the tensor products
\eqnl{
5\otimes 5=1\oplus 10\oplus 14,\quad
4\otimes 4=1\oplus 5\oplus 10,\quad
5\otimes 4=4\oplus 16.}{exb7}
With $10=[0,2]$, $14=[2,0]$ and $16=[1,1]$, we obtain for the Casimir
operators~\refs{ward35}:
\eqnl{
c_5=8,\quad
c_4=5,\quad
c_{10}=12,\quad
c_{14}=20,\quad
c_{16}=15.}{exb9}
Together with~\refs{exb7}, \refs{ward11} implies that
\eqnl{
\chi_{10}=\chi_{4}^2-\chi_{5}-1,\quad
\chi_{14}=\chi_{5}^2-\chi_{4}^2+\chi_{5},\quad
\chi_{16}=\chi_{5}\chi_{4}-\chi_{4},}{exb11}
which leads to the Ward identities
\begin{eqnarray}
\left(u(4\partial_v^2 - 2\partial_u^2 - 4\partial_u+6) + v(5\partial_v -
\partial_u\partial_v) - 8\partial_u\right) Z_\Spf(u,v) & = & 0,\label{exb15a}\\
\left(u(5\partial_v - \partial_u\partial_v) + v(2\partial_u - \partial_v^2+6)
- 5\partial_v\right) Z_\Spf(u,v) & = & 0.\label{exb15b}
\end{eqnarray}
Since the center of Spin$(5)$ is smaller than the center of $SU(3)$ these
differential equations are more complicated than the corresponding $SU(3)$
equations in \refs{exa11} and \refs{exa13}. The characteristics of the second
(generally hyperbolic) equation are given by $u=\mbox{const.}$ and
$\frac{u}{v}=\mbox{const.}$, respectively. These families of lines
coincide for $u=0$ (where \refs{exb15b} is parabolic), and we may solve
the characteristic problem given a solution of both equations for $u=0$.

The restriction of the Ward identities to the $v$-axis
\eqnl{
\big(8\partial_u-v(5\partial_v-\partial_u\partial_v)\big)Z_\Spf\big\vert_{u=0}=0,\qquad
\big(5\partial_v-v(2\partial_u-\partial_v^2+6)\big) Z_\Spf\big\vert_{u=0}=0,}{exb17}
can be solved if we differentiate the second equation with respect to $v$
and use the two relations \refs{exb17} to eliminate the $u$-derivatives at
$u=0$. We find the following ordinary differential equation for $Z_\Spf(0,v)$,
\eqnl{
\left(v^2\partial_v^3 + 13v\partial_v^2 - 16v^2\partial_v + 35\partial_v - 48 v\right) 
Z_\Spf(0,v) = 0.}{exb19}
Since $Z_\Spf$ is an even function in $v$ we set $Z_\Spf(0,v)=Z_0\big(x\;\!{=}\;\!
v^2\big)$ and obtain the simple equation
\eqnl{
\left(x^2\partial_x^3 + 8x\partial_x^2 - 4x\partial_x + 12\partial_x - 6\right)Z_0(x) = 0.}{exb21}
The solution is a hypergeometric function ${_1F_2}\big[{3/2\atop 3,\,4}\big|4x\big]$ 
so that
\eqnl{
Z_\Spf(0,v) = \int \dmur(g) \exp\big(v\chi_4(g)\big) = {_1F_2}\big[\textstyle{3/2\atop 3,\,4}\displaystyle\big|4v^2\big].}{exb23}
Plugging this into the second Ward identity \refs{exb15b} we obtain a recursive solution
\eqnl{
Z_\Spf(u,v) = \sum_{n=0}^{\infty} \frac{Z_n(x=v^2)}{n!}\,u^n,}{exb16}
where
\begin{eqnarray}
Z_1(x) & = & \big(6 \pa_x + 2x \pa^2_x - 3\big) Z_0(x),\nonumber \\
Z_n(x) & = & \big[(5+n)\pa_x + 2x\pa^2_x - 3\big] Z_{n-1}(x) - 5(n-1)\pa_x Z_{n-2}(x)
\mtxt{for}n\geq 2.
\end{eqnarray}

As a special case, let us consider the Spin$(5)$-Ward identities on the $u$-axis,
\eqnl{
\big(u(2\partial_v^2 - \partial_u^2 - 2\partial_u + 3) - 4\partial_u\big)
Z_\Spf\big\vert_{v=0} = 0,\;\;
\big(u(5\partial_v - \partial_u\partial_v) - 5\partial_v\big)
Z_\Spf\big\vert_{v=0} = 0.}{exb25}
We differentiate the first equation with respect to $u$ and obtain
\eqnl{
\left(2\partial_v^2 - 5\partial_u^2 - 2\partial_u + 3 + u(2\partial_u\partial_v^2
- \partial_u^3 - 2\partial_u^2 + 3\partial_u)\right)Z_\Spf\big\vert_{v=0} = 0.}{exb27}
Eq.~\refs{exb25} together with the $v$-derivative of \refs{exb15b} at $v=0$ can
be used to eliminate the $v$-derivatives,
\eqnl{
\left(u^2\partial_u^3+(10u-3u^2)\partial_u^2+(20-12u-13u^2)\partial_u
-30u+15u^2\right)Z_\Spf(u,0)=0.}{exb29}
Very probably this cannot be converted into a differential equation for a generalized hypergeometric series. The differential equation is solved by the power series 
\eqnl{
Z_\Spf(u,0) = 1 + u^2 + \sum_{n=3}^\infty\frac{b_n}{n!}\,u^n}{exb31}
provided that the coefficients satisfy the recursion relation
\eqnl{
b_{n+1}=\frac{1}{(n+4)(n+5)}\Big(
3n(n+3)b_n + n(13n+17)b_{n-1} - 15n(n-1)b_{n-2}\Big),\;\; n\geq 2.}{exb33}
Together with $b_0=1$, $b_1=0$ and $b_2=2$, this determines all coefficients in
the expansion \refs{exb31}.
%%%%%%%%%%%%%%%%%%%
\subsection{The group \texorpdfstring{$G_2$}{G\_2}}\label{groupg2}
For this group, the Cartan matrix is
\eqnl{
K_{G_2} = \pmatrix{2&-1\cr -3&2}.}{exg1}
Hence, $\al_1$ is a short root and $\al_2$ a long root, $\al_1^2=2$ and $\al_2^2=6$.
The quadratic Casimir of the representation with highest weight $\mu=[n_1,n_2]$ 
reads
\eqnl{
c_\mu = 2n_1^2+6n_2^2+6n_1n_2+10n_1+18n_2.}{exg3}
The first fundamental representation $7=[1,0]$ coincides with the subrepresentation
of the 8-dimensional spinor representation of Spin$(7)$ leaving an arbitrary
spinor fixed, and the second is just the adjoint representation $14=[0,1]$. The
center of $G_2$ is trivial and we expect no symmetries of the
generating function.

For the Ward identity we need the tensor products
\begin{eqnarray}
7\otimes 7 & = & 1\oplus 7\oplus 14\oplus 27,\nonumber\\
7\otimes 14 & = & 7\oplus 27\oplus 64,\nonumber \\
14\otimes 14 & = & 1\oplus 14\oplus 27\oplus 77\oplus 77',\nonumber\\
7\otimes 27 & = & 7\oplus 14\oplus 27\oplus 64\oplus 77'.\label{exg5}
\end{eqnarray}
We need the last product in order to express the characters
$\chi_\lam$ as functions of the fundamental characters. Note that two
irreducible representations of dimension $77$ appear in the decompositions.
We identify $27=[2,0]$, $64=[1,1]$, $77=[0,2]$ and $77'=[3,0]$ so that the 
quadratic Casimir operators \refs{ward35} take the following values:
\eqnl{
c_7=12,\quad
c_{14}=24,\quad
c_{27}=28,\quad
c_{64}=42,\quad
c_{77}=60,\quad
c_{77'}=48.}{exg7}
Furthermore, we use
\begin{eqnarray}
\chi_{27} & = & \chi_{7}^2 - \chi_{7} - \chi_{14} - 1,\nonumber\\
\chi_{64} & = & \chi_{7}\chi_{14} - \chi_{7}^2 + \chi_{14} + 1,\nonumber\\
%\chi_{30} & = & \chi_{7}^3 - \chi_{7}^2 - 2\chi_{7}\chi_{14} - \chi_{7} - \chi_{14},
%\label{exg9}\\
\chi_{77} & = & -\chi_{7}^3 + \chi_{14}^2 + 2\chi_{7}\chi_{14} + 2\chi_{7}
+ \chi_{14},\label{exg9}\\
\chi_{77'} & = & \chi_7^3 - \chi_7^ 2 - 2\chi_7\chi_{14} - \chi_7 - \chi_{14}\nonumber
\end{eqnarray}
to derive the following $G_2$-Ward identities 
\begin{eqnarray}
0 & = & \Big(u(-2\partial_u^2 + 8\partial_u + 2\partial_v + 14)\nonumber \\
& & \hskip2mm {}+ v(7\partial_u^2 - 3\partial_u\partial_v + 8\partial_u - 7\partial_v-7)
- 12\partial_u \Big) Z_{G_2}(u,v),\label{exg11}\\
0 & = & \Big(u(7\partial_u^2 - 3\partial_u\partial_v + 8\partial_u - 7\partial_v - 7)\nonumber\\
& & \hskip2mm {}+v(6\partial_u^3 + 10\partial_u^2 - 6\partial_v^2 - 12\partial_u\partial_v - 22\partial_u - 4\partial_v+14) - 24\partial_v \Big)Z_{G_2}(u,v)\label{exg13}
\end{eqnarray}
for the generating function
\eqnl{
Z_{G_2}(u,v) = \int\dmur(g) \exp\big(u\chi_7(g) + v\chi_{14}(g)\big).}{exg10}
On the $u$-axis these equations simplify to
\begin{eqnarray}
0 & = & \left(u(-\partial_u^2 + 4\partial_u + \partial_v + 7) - 6\partial_u \right) Z_{G_2}(u,v)\vert_{v=0}, \label{exg15}\\
0 & = & \left(u(7\partial_u^2 - 3\partial_u\partial_v + 8\partial_u - 7\partial_v - 7) - 24\partial_v \right)Z_{G_2}(u,v)\vert_{v=0}.\label{exg17}
\end{eqnarray}
We solve the first equation and its $u$-derivative for the $v$-derivatives 
at $v=0$ occurring in the second equation and end up with the third order
differential equation
\eqnl{
\left(u^2\partial_u^3 + (14u-4u^2)\partial_u^2 + (42-18u-19u^2)\partial_u - 14u^2 - 56 u\right)
Z_{G_2}(u,0)=0.}{exg19}
Again this is probably not related to a generalized hypergeometric
series. It may be solved in terms of a series expansion 
\eqnl{
Z_{G_2}(u,0) = 1 + u^2 + \sum_{n\geq 3} \frac{g_n}{n!} u^n}{exg21}
provided that the coefficients satisfy the recursion relation (for $n\geq 2$)
\eqnl{
g_{n+1} = \frac{1}{(n+6)(n+7)}\Big(2n(2n+7)g_n + n(19n+37) g_{n-1}
+ 14n(n-1)g_{n-2}\Big),}{exg23}
together with $g_0=1$, $g_1=0$, $g_2=2$. These coefficients are related to the
triangulations of $n$-gones with inner vertices with valences $\geq 6$~\cite{kupferberg}.

Starting from $Z_{G_2}(u,0)=\tilde{Z}_0(u)$, we can now solve the corresponding
characteristic problem for the first Ward identity~\refs{exg11} (analogously
to the Ward identity for Spin$(5)$) by means of an expansion
\eqnl{Z_{G_2}(u,v) = \sum_{n=0}^\infty \tilde{Z}_n(u)v^n}{exg25}
with
\begin{eqnarray}
u \tilde{Z}_1(u) & = & \big( u\pa_u^2 - 4u\pa_u + 6\pa_u - 7u\big) \tilde{Z}_0(u),\nonumber\\
2u \tilde{Z}_n(u) & = & (1-n)\big( 7\pa_u^2 + 8\pa_u - 7\big) \tilde{Z}_{n-2}(u) \nonumber\\
& & \;{}+ \big( 2u\pa_u^2 - 8u\pa_u + 3(n+3)\pa_u - 14u + 7n - 7\big) \tilde{Z}_{n-1}(u)
\label{exg27}
\end{eqnarray}
for all $n\geq 2$. With the help of \refs{exg23} and \refs{exg27}, one can reproduce the moments
given in the appendix (section~\ref{append2}) recursively.

%%%%%%%%%%%%%%%%%%%%%%%
\section{Results for \texorpdfstring{$SU(4)$}{SU(4)}} \label{rk3}
In this section, we will derive a solution to the Ward identities of
the rank~$3$ group $SU(4)$ in a certain range of the parameter values.
For $SU(4)$, the quadratic Casimir of the representation with highest
weight $\mu=[n_1,n_2,n_3]$ is given by
\eqnl{
c_\mu = \frac{1}{4}(3n_1^2 + 4n_2^2 + 3n_3^3 + 4n_1n_2 + 2n_1n_3 + 4n_2n_3)
+ 3n_1 + 4n_2 + 3n_3,}{suv1}
and the fundamental representations $4$, $6$, $\bar 4$ with highest weights
$\mu_1\equiv[1,0,0]$, $\mu_2\equiv[0,1,0]$, and $\mu_3\equiv[0,0,1]$ have Casimirs
\eqnl{
c_4 = c_{\bar 4} = \frac{15}{4}\mtxt{and}c_6 = 5.}{suv3}
The real representation $6$ coincides with the vector representation of $SO(6)$, and
$\bar{4}$ is complex conjugated to the defining representation $4$; the latter two can
be identified with the complex fundamental spinor representations of Spin$(6)$. Their
tensor products can be decomposed according to
\begin{eqnarray}
4\otimes 4 = 6\oplus 10,\quad
6\otimes 6 \!\!\!& = &\!\!\! 1\oplus 20\oplus 15,\quad
\bar{4}\otimes\bar 4 = 6\oplus \overline{10},\nonumber\\
4\otimes\bar 4 = 1\oplus 15,\quad
4\otimes 6 \!\!\!& = &\!\!\!\bar 4 \oplus 20',\quad
\bar{4}\otimes 6 = 4\oplus\overline{20}',\label{suv5}
\end{eqnarray}
where again we denoted the representations by their dimensions,
\eqnn{
10=[2,0,0],\;\overline{10}=[0,0,2],\; 15=[1,0,1],\;
20=[0,2,0],\; 20'=[1,1,0],\;\overline{20}'=[0,1,1].}
The representations $15$ and $20$ are real and $\bar{4}$, $\overline{20}'$ are
complex conjugate to $4$, $20'$. From \refs{suv5}, we find
\begin{eqnarray*}
\chi_{10}=\chi_4^2-\chi_6,\quad
\chi_{15}=\chi_4\bar{\chi}_{4}-1,\quad
\chi_{20}=\chi_6^ 2-\chi_4\bar{\chi}_{4},\quad
\chi_{20'}=\chi_4\chi_6-\bar{\chi}_{4}.
\end{eqnarray*}
These data enter the Ward identities \refs{ward23} for the generating function
\eqnl{
Z_{SU(4)}(u,v,w)=\int d\mu_{\rm red}\ e^{u\chi_4+v\bar\chi_4+w\chi_6},}{suv9}
which is center-symmetric, $Z_{SU(4)}\left(iu,-iv,-w\right)=Z_{SU(4)}(u,v,w)$.
They take the form
\begin{eqnarray}
\!\!0 &\!\! =\! & \hskip-2mm\left\{15\partial_u + u(3\partial_u^2 - 8\partial_w)
+ 2w(\partial_u\partial_w - 6\partial_v) + v(\partial_u\partial_v - 16)\right\}Z_{SU(4)}(u,v,w),\label{suv11a}\\
\!\!0 &\!\! =\! & \hskip-2mm\left\{10\partial_w + u(\partial_u\partial_w - 6\partial_v)
+ 2w(\partial_w^2 - 2\partial_u\partial_v - 4)
+ v(\partial_v\partial_w - 6\partial_u)\right\}Z_{SU(4)}(u,v,w),\label{suv11b}\\
\!\!0 &\!\! =\! & \hskip-2mm\left\{15\partial_v + v(3\partial_v^2 - 8\partial_w)
+ 2w(\partial_v\partial_w - 6\partial_u)
+ u(\partial_u\partial_v - 16)\right\}Z_{SU(4)}(u,v,w).\label{suv11c}
\end{eqnarray}
In order to find an explicit solution for arbitrary products of 4-dimensional
representations (i.e., on the diagonal with $u=v$ and $w=0$), we proceed
analogously to section~\ref{groupa2}: We act with the operator
$(u\partial_u+v\partial_v+9)$ on the first and last equations and use
\refs{suv11b} to eliminate the $w$-derivative in the resulting
differential equations. At $w=0$, we obtain
\begin{eqnarray}
& & \left(3u^2\partial_u^3 + 4uv\pa_u^2\pa_v + v^2\pa_u\pa_v^2 + 45u\pa_u^2 +
25v\partial_u\partial_v\right)Z_{SU(4)}(u,v,0)\nonumber \\
& & \:\,{}+ \left((135-64uv)\pa_u - (16v^2+48u^2)\pa_v - 160 v\right)Z_{SU(4)}(u,v,0)
= 0,\nonumber\\
& & \left(3v^2\pa_v^3 + 4uv\pa_u\pa_v^2 + u^2\pa_u^2\pa_v + 45v\pa_v^2 +
25u\pa_u\pa_v\right)Z_{SU(4)}(u,v,0)\nonumber\\
& & \:\,{}+ \left((135-64uv)\pa_v - (16u^2+48v^2)\pa_u - 160u\right)Z_{SU(4)}(u,v,0)
= 0.\label{suv13}
\end{eqnarray}
Acting with the operator $(u\pa_u+2u\pa_v+7)$ on the first equation
in \refs{suv13} and evaluating the result at $u=v=t$ and $w=0$, we obtain
the following differential equation for $Z(t,t,0)$,
\eqnl{\Big(t^3\frac{d^4}{dt^4} + 24\:\!t^2\frac{d^3}{dt^3} + (165-64\:\!t^2)t
 \frac{d^2}{dt^2} + 9(35-64t^2)\frac{d}{dt} - 960\:\!t\:\!\Big) Z_{SU(4)} = 0,}{suv15}
where the derivatives with respect to $t$ are given by expressions analogous
to~\refs{exa27}. Restricting $Z_{SU(4)}$ to the diagonal breaks the $\Z_4$ center
symmetry down to $\Z_2$; we identify \refs{suv15} in terms of the $\Z_2$-invariant
coordinate $y=16\:\! t^2$,
\eqnl{
\big(y^4\pa_y^4 + 15y^3\pa_y^3 + (60y^2-y^3)\pa_y^2 + (60y-5y^2)\pa_y - \ft{15}{4}y\big)Z_{SU(4)}=0,}{suv17}
as the defining equation~\refs{einf16} for the hypergeometric function
\eqnl{Z_{SU(4)}(t,t,0) = \int\dmur(g) \exp\big(t(\tr g + \tr g^\dagger)\big)
= {}_2F_3\Big[{3/2,5/2\atop 3,4,5}\Big|16\:\!t^2\Big].}{suv19}
This result proves the conjecture in~\cite{carlsson} (which is based on numerical
observations). As a by-product, this also leads to the remarkable identity
\eqnl{
\sum_{n\in\Z}\det \pmatrix{ I_n & I_{n+1} & I_{n+2} & I_{n+3}\cr
I_{n-1}  & I_n & I_{n+1} & I_{n+2}\cr
I_{n-2} & I_{n-1} & I_n & I_{n+1}\cr
I_{n-3} & I_{n-2} & I_{n-1} & I_n}\!\!\!\:(2t)\; =\;
{}_2F_3\Big[{3/2,5/2\atop 3,4,5}\Big|16\:\!t^2\Big],}{suv21}
cf.~eq.~\refs{einf11} for $u=v=t$, relating generalized hypergeometric
functions and determinants of Bessel functions.
%%%%%%%%%%%%%%%%%%
\section{On the reduced Haar measure}\label{redHM}
In this section, we will describe an alternative approach to Ward identities
for the generating function~$Z_G$ based on a factorization of the reduced Haar
measure on the maximal Abelian torus in~$G$. Tangent vectors to this torus are
linear combinations $H_\varphi=\sum_p \varphi_p H_p$ of the Cartan generators
$H_p$. The reduced Haar measure $\dmur=\rho_{\rm red}\,d^r\varphi$ on the maximal 
abelian torus has the product representation~\cite{Fulton}
\eqnl{
\rho_{\rm red} \left(e^{iH_\varphi}\right) \propto 
\prod_{\al >0} 4\sin^2\left(\ha\,\al(H_\varphi)\right) =
\prod_{\mbm\in\Phi^+}4\sin^2\left(\ha(\mbm,K\mbvarphi)\right)}{rhaar1}
with one factor for every positive root~$\al$. The $\ha\big(\hbox{dim}(G)-
\hbox{rank}(G)\big)$ positive roots are linear combination of the
simple roots,
\eqnl{
\alpha=m_1\al_{(1)}+m_2\al_{(2)}+\dots+ m_r\al_{(r)},\quad m_i\in \N_0,}{rhaar3}
and the range $\Phi^+$ for $\mbm=(m_1,\ldots,m_r)^t$ in~\refs{rhaar1} is chosen
in such a way that it parametrizes all positive roots. We may take the square root
of the density~\cite{Fulton},
\eqnl{
\rho_{\rm red} \left(e^{iH_\varphi}\right) \propto \vert\Delta\vert^2,\quad
\Delta = \prod_{\mbm\in\Phi^+} 2i\sin\left(\ha (\mbm,K\mbvarphi)\right)
= \sum_{w\in W} \hbox{sign}(w)e^{iw(\rho)(H_\varphi)},}{rhaar5}
where the sum runs over the Weyl orbit $W$ of the Weyl vector $\rho$ introduced
in \refs{ward31}. Since the Weyl orbit of $\rho$ contains $|W|$ elements the product
representation is preferable for large groups.\footnote{For example for $SU(N)$
the sum has $N!$ terms, whereas the product has only $\ha N(N-1)$
factors.} But it is evident from the sum representation that $\Delta$ changes
sign under Weyl reflections.

The density $\rho_{\rm red}\propto\Delta\bar\Delta$ of the reduced Haar
measure is a Weyl-invariant function on the maximal Abelian torus
and hence a function of the fundamental characters. From
\refs{rhaar1} we see that it actually is a \emph{polynomial} of the fundamental 
characters. In contrast, $\Delta$ is not Weyl-invariant and hence cannot be
written as function of the characters.

Less obvious is the observation that $\Delta$ is related to the Jacobian
of the transformation $\mbvarphi\mapsto\mbchi(\mbvarphi)$ from the angular
variables to the fundamental characters,
\eqnl{
J(\mbchi)\equiv \big\vert \det\left(\frac{\partial\mbchi}{\partial\mbvarphi}\right)
\big\vert\propto \vert\Delta\vert,\mtxt{such that}
\dmur=J^2 d^r\varphi=J(\mbchi)d^r\chi.}{rhaar7}
This mapping is one-to-one on the fundamental domain $\cF$ of the action
of the Weyl group on the maximal Abelian torus. This is the closed connected
region containing $\mbchi=0$ in which $\Delta\geq 0$. We have no proof of the
\emph{conjecture} \refs{rhaar7} for all compact simple groups, but have checked
it for the groups $SU(3)$, Spin$(5)$ and $G_2$ considered
in the following sections as well as for $SU(2)$ and $SU(4)$.

Based on this conjecture we may derive alternative Ward identities from
\eqnl{
\int_\cF  d^r\chi\, \frac{\partial}{\partial\chi_p}\left(J^{3}(\mbchi) F(\mbchi)\right)
= 0,}{rhaar9}
where use was made of the fact that the Jacobian vanishes on the
boundary of the fundamental domain. This leads to the general and simple
looking Ward identities
\eqnl{
0 = \int_\cF \dmur \left(\frac{3}{2}\frac{\partial J^2 }{\partial \chi_p}F +
J^2\frac{\partial F}{\partial\chi_p}\right),\qquad p=1,\ldots,r}{rhaar11}
for any regular function $F=F(\mbchi)$ on the fundamental domain.
In particular they imply the following differential identities for the
generating function $Z_G(\mbu)$:
\eqnl{
\left(\frac{3}{2}\frac{\partial J^2}{\partial \chi_p}(\mbfgr{\partial}) + u_p J^2(\mbfgr{\partial})\right)Z_G(\mbu) = 0,\qquad p=1,\dots,r.}{rhaar13}
These should be compared with the geometric Ward identities~\refs{ward23}.
In reference~\cite{wozar} by one of the authors, both types of
identities were applied to calculate effective Polyakov loop dynamics of
$SU(3)$ Yang-Mills theories on the lattice. The geometric Ward identities
are usually simpler but not necessarily favored in computer simulations.

Ultimately, the two systems of linear partial differential equations~%
\refs{rhaar13} and \refs{ward23} must be equivalent; but a proof is not
straightforward at all. For example, for $SU(3)$ the equations \refs{rhaar13}
are $4$th order differential equation whereas \refs{ward23} are of
$2$nd order.

Now we apply the general result \refs{rhaar13} to all simple compact
simply-connected groups of rank $2$. As mentioned above, for these group
the conjecture that the density of the reduced Haar measure is proportional
to the square of the Jacobian $J$ of the transformation $\mbvarphi\mapsto\mbchi$
can be checked explicitly. The Jacobians for the three groups are computed in
the following subsections.

\subsection{The group \texorpdfstring{$SU(3)$}{SU(3)}}
\enlargethispage*{3mm}
As in section \ref{groupa2}, $[1,0]$ denotes the defining representation $3$
and $[0,1]$ its complex conjugate $\bar 3$. The reduced Haar measure reads
\eqnl{
\dmur = \frac{1}{6\pi^2} J^2\,d\varphi_1d\varphi_2 =
\frac{1}{6\pi^2} J(\mbchi)\,d\chi_3 d\chi_{\bar 3},}{rha1}
with $J^2\propto \rho_{\rm red}$ from \refs{rhaar1}. As a function of the
fundamental characters the Weyl-invariant and center-symmetric $J^2$ reads
\eqnl{
J^2 = 27 + \chi_3^3 + \chi_{\bar 3}^3 - \frac{1}{4}\left(9 +
\chi_3\chi_{\bar 3}\right)^2.}{rha3}
It can be easily verified that its positive square root $J$ indeed coincides
with the Jacobian of the map $(\varphi_1,\varphi_2)\mapsto (\chi_3,\chi_{\bar 3})$.
\\[3mm]
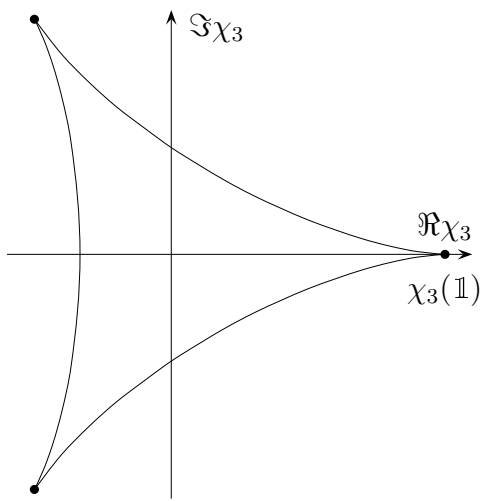
\begin{minipage}[t]{6.8cm}
\psset{unit=1.2cm,linewidth=0.3pt,arrowsize=1.4mm,dotsize=1.2mm}
\begin{pspicture}(-2,2.8)(3.3,-2.7)
\psline{->}(-1.8,0)(3.3,0)
\psline{->}(0,-2.7)(0,2.7)
\rput(.5,2.5){$\Im \chi_3$}
\rput(3,0.3){$\Re \chi_3$}
\psdot(3,0)
\psdot(-1.5,2.598)
\psdot(-1.5,-2.598)
\rput(3,-.4){$\chi_3(\id)$}
\pscurve
(-1.5,2.60)(-1.2,2.21)(-0.9,1.90)(-0.6,1.63)(-0.3,1.40)
(0.0,1.18)(0.3,0.99)(0.6,0.81)(0.9,0.65)(1.2,0.51)
(1.5,0.38)(1.8,0.27)(2.1,0.17)(2.4,0.09)(2.7,0.03)
(3,0.00)
\pscurve
(-1.5,-2.60)(-1.2,-2.21)(-0.9,-1.90)(-0.6,-1.63)(-0.3,-1.40)
(0.0,-1.18)(0.3,-0.99)(0.6,-0.81)(0.9,-0.65)(1.2,-0.51)
(1.5,-0.38)(1.8,-0.27)(2.1,-0.17)(2.4,-0.09)(2.7,-0.03)
(3,-0.00)
\pscurve
(-1.50,2.60)(-1.40,2.38)(-1.30,2.10)(-1.20,1.74)(-1.10,1.25)
(-1.00,0.00)(-1.10,-1.25)(-1.20,-1.74)(-1.30,-2.10)(-1.40,-2.38)(-1.50,-2.60)
\end{pspicture}
\end{minipage}
\,\hfill\hfill
\begin{minipage}[b]{8cm}
The Jacobian vanishes for
\begin{eqnarray*}
y^2=-(9+12x+x^2)\pm 2(2x+3)^{3/2},
\end{eqnarray*}
where $x=\Re(\chi_3)$ and $y=\Im(\chi_3)$ are the real and imaginary parts
of $\chi_3$. The fundamental domain inside the triangularly shaped region
is symmetric under $\Z_3$ center transformations which rotate $\chi_3$
by multiples of $e^{2\pi i/3}$. Its corners are the values of $(\Re\chi_3,
\Im\chi_3)$ at the center elements. Here, the identities \refs{rhaar13} for the
generating function take the form
\end{minipage}
\begin{eqnarray}
0 & = & \Big( 3\big(6\partial_u^2 - 9\partial_v - \partial_u\partial^2_v\big)
+ u\big(27 + 4\partial_u^3 + 4\partial_v^3 - 18\partial_u\partial_v - \partial_u^2\partial_v^2\big)\Big) Z_{SU(3)},
\label{rha5}\\
0 & = & \Big( 3\big(6\partial_v^2 - 9\partial_u - \partial_v\partial_u^2\big)
+ v\big(27 + 4\partial_u^3 + 4\partial_v^3 - 18\partial_u\partial_v -  \partial^2_u\partial_v^2\big)\Big)Z_{SU(3)}.
\label{rha7}
\end{eqnarray}
In contrast to the geometric identities (\ref{exa11}, \ref{exa13})
these are $4$th order differential equations.
%%%%%%%%%%%%%%%%
\subsection{The group \texorpdfstring{Spin$(5)$}{Spin(5)}}
With the conventions used in section \ref{groupb2}, $[1,0]=5$ is the defining
representation and $[0,1]=4$ denotes the spin representation. The reduced Haar
measure reads
\eqnl{
\dmur\propto J^2 d\varphi_1 d\varphi_2 = J d\chi_5 d\chi_4}{rhb1}
with Jacobian $J$ such that
\eqnl{
J^2=(3+\chi_5-2\chi_4)(3+\chi_5+2\chi_4)(4-4\chi_5+\chi_4^2).}{rhb3}
\\
\begin{minipage}[t]{6.8cm}
\psset{unit=.8cm,linewidth=0.3pt,arrowsize=1.4mm,dotsize=1.2mm}
\begin{pspicture}(-3.4,-4.1)(5.2,4.3)
\psline{->}(-3.2,0)(5.2,0)
\psline{->}(0,-4.1)(0,4.2)
\psline[showpoints=true](5,4)(-3,0)(5,-4)
\rput(5,.4){$\chi_5$}
\rput(.5,3.8){$\chi_4$}
\small
\rput(5,3.3){$\mbchi(\id)$}
\rput(-2.8,.8){$(-3,0)$}
\rput(5,-3.3){$\mbchi(z)$}
\pscurve
(5.00,-4.00)(4.60,-3.79)(4.20,-3.58)(3.80,-3.35)(3.40,-3.10)
(3.00,-2.83)(2.60,-2.53)(2.20,-2.19)(1.80,-1.79)(1.40,-1.26)
(1.00,0.00)(1.40,1.26)(1.80,1.79)(2.20,2.19)(2.60,2.53)
(3.00,2.83)(3.40,3.10)(3.80,3.35)(4.20,3.58)(4.60,3.79)
(5.00,4.00)
\end{pspicture}
\end{minipage}
\,\hfill\hfill
\begin{minipage}[b]{8cm}
The zero locus of the Jacobian is given by
\begin{eqnarray*}
 2y = \pm (x+3),\quad
  y = \pm\sqrt{x-1},
\end{eqnarray*}
where we abbreviated $x=\chi_5$ and $y=\chi_4$.
The fundamental domain inside the triangularly shaped region is invariant
under the $\Z_2$ center symmetry flipping the sign of $\chi_4$. Two of the
corners, $(\chi_5(\id),\chi_4(\id))=(5,4)$ and $(\chi_5(z),\chi_4(z))=(5,-4)$,
respectively, are located at the values of the characters at the center elements.
The identities \refs{rhaar13} for the generating function take the form
\end{minipage}
\begin{eqnarray}
\hskip-5mm 0\hskip-2mm & = & \hskip-3mm \Big( u\big( (3 + \partial_u)^2 -
4\partial^2_v\big)\big(4 - 4\partial_u + \partial_v^2\big) +
3\big(\partial_u\partial_v^2 - 6\partial_u^2 + 11\partial_v^2 - 20\partial_u
- 6\big)\Big)Z_\Spf, \label{rhb5}\\
\hskip-5mm 0\hskip-2mm & = & \hskip-3mm \Big( v\big( (3 + \partial_u)^2 -
4\partial^2_v\big)\big(4 - 4\partial_u + \partial_v^2\big) +
3\big(\partial_u^2\partial_v - 8\partial_v^3 + 22\partial_u\partial_v
- 7\partial_v\big)\Big)Z_\Spf.\label{rhb7}
\end{eqnarray}
Again, these are $4$th order differential equations.
%%%%%%%%%%%%%%%%%%%%
\subsection{The group \texorpdfstring{$G_2$}{G\_2}}
With the conventions of section \ref{groupg2}, $7=[1,0]$ denotes the
$7$-dimensional representation and $14=[0,1]$ the adjoint representation.
The density $J^2$ of the reduced Haar measure
\eqnl{
\dmur\propto J^2 d\varphi_1 d\varphi_2 = J d\chi_5 d\chi_4}{rhg1}
is a quintic polynomial in $\chi_7$ and a cubic polynomial in $\chi_{14}$, 
\eqnl{
J^2 = \left(4\chi_7^3-\chi_7^2-2\chi_7-10\chi_7\chi_{14}
+ 7 - 10\chi_{14} - \chi_{14}^2\right) \left(7 - \chi_7^2 - 2\chi_7 +
4\chi_{14}\right).}{rhg3}
Since the center of $G_2$ is trivial, this polynomial shows no symmetries
at all. Nevertheless, it is possible to characterize the fundamental domain
for the exceptional group $G_2$ explicitly.
\vskip2.4mm
\begin{minipage}[t]{6.8cm}
\psset{xunit=.7cm,yunit=.4cm,linewidth=.3pt,arrowsize=1.4mm,dotsize=1.2mm}
\begin{pspicture}(-3,-3)(7.2,14.2)
\psline{->}(-2.5,0)(7.2,0)
\psline{->}(0,-2.2)(0,14.1)
\rput(7,.7){$\chi_5$}
\rput(.7,13.5){$\chi_{14}$}
\small
\rput(-1,-2.8){$(-1,-2)$}
\rput(6,14){$\mbchi(\id)$}
\rput(-2,5.6){$(-2,5)$}
\psdot(7,14)
\psdot(-1,-2)
\psdot(-2,5)
\pscurve
(-1.00,-2.00)(-0.60,-1.96)(-0.20,-1.84)(0.20,-1.64)(0.60,-1.36)
(1.00,-1.00)(1.40,-0.56)(1.80,-0.04)(2.20,0.56)(2.60,1.24)
(3.00,2.00)(3.40,2.84)(3.80,3.76)(4.20,4.76)(4.60,5.84)
(5.00,7.00)(5.40,8.24)(5.80,9.56)(6.20,10.96)(6.60,12.44)
(7.00,14.00)
\pscurve
(-2.00,5.00)(-1.50,3.21)(-1.00,2.00)(-0.50,1.17)(0.00,0.66)
(0.50,0.41)(1.00,0.39)(1.50,0.60)(2.00,1.00)(2.50,1.59)
(3.00,2.36)(3.50,3.30)(4.00,4.39)(4.50,5.64)(5.00,7.04)
(5.50,8.58)(6.00,10.25)(6.50,12.06)(7.00,14.00)
\pscurve
(-2.00,5.00)(-1.80,3.82)(-1.60,2.49)(-1.40,1.07)(-1.20,-0.43)
(-1.00,-2.00)
\end{pspicture}
\end{minipage}
\,\hfill\hfill
\begin{minipage}[b]{7.2cm}
The zero locus of the Jacobian is given by
\begin{eqnarray*}
y & = & \frac{1}{4}(x+1)^2 - 2,\\
y & = & -5(x+1)\pm 2(x+2)^{3/2},
\end{eqnarray*}
where we introduced $x=\chi_7$ and $y=\chi_{14}$. The fundamental domain
is the region bounded by the three curves defined by the above equations.
The upper right corner is located at the characters of the unit element, $(\chi_7,\chi_{14})=(7,14)$. In this case, the identity \refs{rhaar13}
for the generating function reads
%\hfill
%\begin{minipage}[b]{7.8cm}
\end{minipage}
\begin{eqnarray}
0 & = & \Big(u\big( 4\partial_u^3 - \partial_u^2 - 2\partial_u - 10\partial_u\partial_v
+ 7 - 10\partial_v - \partial_v^2\big)\big(7 - \partial_u^2 - 2\partial_u + 4\partial_v
\big)\nonumber\\
& & \;{}+ 3\big(3 - 29\partial_u + 13\partial_u^3 + 13\partial_u^2 - 38\partial_u\partial_v - 21 - 47\partial_v + \partial_u^2\partial_v - 6\partial_v^2 \big)\Big)Z_{G_2},\label{rhg5}\\
0 & = & \Big(v\big( 4\partial_u^3 - \partial_u^2 - 2\partial_u - 10\partial_u\partial_v + 7 - 10\partial_v - \partial_v^2\big)\big(7-\partial_u^2-2\partial_u+4\partial_v
\big)\nonumber\\
& & \;{}+ \big(3 - 29\partial_u + 13\partial_u^3 + 13\partial_u^2 - 38\partial_u\partial_v - 21
- 47\partial_v + \partial_u^2\partial_v - 6\partial_v^2\big)\Big)Z_{G_2}.\label{rhg7}
\end{eqnarray}
Since $J^2$ is quintic we arrive at complicated $5$th order
linear partial differential equations which should be compared 
with the equivalent geometric Ward identites (\ref{exg11}, \ref{exg13}).

%%%%%%%%%%%%%%%%%%%%
\section{Recursion relations for the moments}\label{recrel}
The function $Z_G(\mbu)$ generates the moments
\eqnl{
t_{m_1,\dots,m_r}=
\int \dmur\, \chi_1^{m_1}\cdots \chi_r^{m_r} }{mom1}
by multiple differentiation at $\mbu=\mb0$, see \refs{einf4}.
Due to center symmetry of the Haar measure only center symmetric moments are
nonzero, and this selection rule must be respected by any recursion relation
for the moments.

One may use the Ward identity for $Z_G(\mbu)$ to find such relations. A more
direct derivation takes advantage of \refs{ward15} with $F=\chi_1^{m_1}\cdots
\chi_r^{m_r}$. One finds
\eqngrl{
0 & = & \Big(2c_p - \sum_q (c_p + c_q)m_q\Big)t_{m_1,\dots,m_p+1,\dots m_r}}
{& & \;{}+\sum_{\lam,q}C^\lam_{pq}c_\lam m_q \int\dmur\,\chi_\lam(\mbchi) \chi_1^{m_1}\cdots\chi_q^{m_q-1}\cdots \chi_r^{m_r}.}{mom3}
Clearly, the complexity of these relations increases with the degree
of the polynomials $\chi_\lam$ in the last sum.
Alternatively, we could apply eq.~\refs{rhaar11} to the same function $F$,
with the result
\eqnl{
0 = \int\dmur \left(3\,\frac{\partial J^2(\mbchi)}{\partial\chi_p}\chi_1^{m_1}
\cdots\chi_r^{m_r} + 2m_p J^2(\mbchi)\chi_1^{m_1}\cdots\chi_p^{m_p-1}\cdots
\chi_r^{m_r}\right) .}{mom5}
These relations are based on the conjecture \refs{rhaar7},
in contrast to the `geometric recursion relations' in \refs{mom3}.
%%%%%%%%%%%%%%%%%%%%%%%%%%%%%%%%%%
\paragraph{For the group SU(3):}
For this group the recursion relations \refs{mom3} take the form
\begin{eqnarray}
(8+2m+n)t_{m+1,n}-6m t_{m-1,n+1}-9n t_{m,n-1}&=&0,\label{moma1}\\
(8+2n+m)t_{m,n+1}-6n t_{m+1,n-1}-9m t_{m-1,n}&=&0.\label{moma3}
\end{eqnarray}
Since the moments are symmetric the two identities are equivalent. 
These `geometric identities' are much simpler than the `non-geometric'
relations \refs{mom5}, which for $p=1$ lead to
\eqnn{
 (18+4m)t_{n,m+2} - (3+m)t_{n+2,m+1} - (27+18m)t_{n+1,m} +
 27m t_{n,m-1} + 4mt_{n+3,m-1} = 0.}
By symmetry of the coefficients $t_{mn}$, the relation for $p=2$ is again
equivalent to this recursion formula. The difference of both leads to the
simpler formula
\eqngrl{
0 & = & (4k+6)t_{3k+m+3,m} + (4k-6)t_{3k+m,m+3}}
{& & \;{}+ k\left(27t_{3k+m,m} - 18t_{3k+m+1,m+1} - t_{3k+m+2,m+2}\right).}{moma5}
All recursion relations are compatible with center symmetry which
implies $t_{mn}=0$ unless $m=n$ mod $3$. With the moments
\eqnl{
t_{3m,0}=\frac{2(3m)!}{m!(m+1)!(m+2)!},\quad
t_{3m+1,1}=\frac{6(3m+1)!}{m!(m+1)!(m+3)!}}{moma7}
one can compute all $t_{mn}$ with the recursion relation \refs{moma1}. For example,
for $m=n$, one finds
\eqnl{
t_{mm} = 2\sum_{k=0}^m {2k\choose k}{m\choose k}^2\frac{3k^2+2k+1-2km-m}
{(k+1)^2(k+2)(m-k+1)}.}{moma9}
The moments $t_{mn}$ for small $m$ and $n$ are given in the appendix.
\paragraph{For the group Spin(5):}
For this group the geometric recursion relations \refs{mom3} read
\begin{eqnarray}
(8+2m+n)t_{m+1,n} + (4m-5n) t_{m,n} - 6mt_{m-1,n} - 4m t_{m-1,n+2} & = & 0,\label{momb1}\\
(5+m+n)t_{m,n+1} - 5m t_{m-1,n+1} - 2n t_{m+1,n-1} - 6nt_{m,n-1} & = & 0.\label{momb3}
\end{eqnarray}
These relations are compatible with center symmetry which implies
that $t_{mn}=0$ for odd $n$. With the help of the first or second recursion
relations, one can determine any $t_{mn}$ given $t_{m,0}$. For the first
relation we also need the $t_{0,n}$ which are just the coefficients in the
series expansion of $Z(0,v)$ in \refs{exb23}. The moments $t_{mn}$ for small
$m$ and $n$ can be found in the appendix.
\paragraph{For the group G$_2$:}
For this exceptional group the recursion relation \refs{mom3}
are more involved,
\begin{eqnarray}
0 & = & 2m t_{m-1,n+1} - (12+2m+3n)t_{m+1,n} + (8m-7n)t_{m,n}\nonumber\\
& & \;{}+ 14mt_{m-1,n} + 7n t_{m+2,n-1} + 8nt_{m+1,n-1} - 7nt_{m,n-1},\label{momg1} \\
0 & = & (24+3m+6n)t_{m,n+1} + 7mt_{m-1,n+1} - (7m-12n)t_{m+1,n} - (8m-4n)t_{m,n}\nonumber\\
& & \;{}+ 7mt_{m-1,n} - 6nt_{m+3,n-1}-10n t_{m+2,n-1}+22nt_{m+1,n-1}-14nt_{m,n-1}.\label{momg3}
\end{eqnarray}
For example, one can calculate all $t_{mn}$ from the $t_{m,0}$ and $t_{m,1}$.
The former ones are just the coefficients $g_m$ in \refs{exg23}. The moments
$t_{mn}$ for small $m$ and $n$ are given in the appendix.

\section{Conclusions}
In this paper, we have derived two kinds of Ward identities for the generating
functions for integrals over arbitrary polynomials of fundamental characters.
One is a consequence of the fact that a left-derivative of any function on the
Lie group integrates to zero with the full Haar measure. For a convenient choice
of this function, this left-derivative is a class function so that the vanishing
of the integral reduces to an identity on the maximal Abelian torus. The other
Ward identity derives from an integral of a total derivative of an arbitrary
class function over a certain domain. If one chooses this domain as the region
where the Jacobian of the change of variables from the angles of the maximal
Abelian torus to the fundamental characters is non-negative, one can split powers
of the Jacobian from the arbitrary class function so that the result vanishes
on the boundary of this fundamental domain. This, however, leads to ultimately more
complicated differential equations for the generating functions than the first,
geometric, approach. Both furnish generalizations and structural clarifications
of identities used in the case of $SU(3)$ in an earlier publication~\cite{wozar}
by one of the authors. In this paper, they have been applied to all simple compact
simply connected Lie groups of rank two. Furthermore, they have been used to prove
several conjectures in the literature concerning explicit solutions for $SU(3)$ and
$SU(4)$; beyond that, we derived recursion relations determining all integrals
over polynomials of fundamental characters for the above groups.

The derivation of the second kind of Ward identities is based on a conjecture
concerning the factorization of the reduced Haar measure density $\rho_{\rm red}$
into the square of this Jacobian. This conjecture has been checked explicitly for
several cases under consideration, but so far it lacks a general proof. It might
be interesting to clarify this issue from a group theoretical point of view.

Another obvious open question is the equivalence of our two approaches. Since both
encode information determining the same generating functions it ultimately should
be possible to derive one from the other. Perhaps with a deeper insight into the
group theoretical connection between the approaches it might be possible to obtain
even simpler identities which might be of use, e.g., in lattice gauge theories or
random matrix models. From a mathematical point of view they answer the question
how many invariants there are in a given tensor product of fundamental
representations.

We found it surprising to note that the generating function for powers of the
sum of characters of only the defining and its complex conjugate representation
for $SU(2)$ and $SU(4)$ can be expressed in terms of appropriate generalized
hypergeometric functions. One might speculate that this generalizes to higher
rank $SU(2n)$ as well. Apart from that, this fact for $SU(4)$ leads to a hitherto
unknown relation~\refs{suv21} between generalized hypergeometric functions and
Bessel functions.
Finally, one might note the remarkable fact that all Ward identities (of the
first kind), at least for the Lie groups of rank~$2$, reduce to parabolic
differential equations on the locus where their families of characteristics
coincide. This allows for a recursive integration starting from a solution on
this locus.
\\[3mm]
\textbf{Acknowledgements:} We thank John Harnad and Christian Lang
for discussions as well as Jan Steinhoff and Christian Wozar for a collaboration
at an early stage of our investigations. This project has in part been
supported by the DFG, grant Wi 777/8-2. The explicit group-theoretical
calculations have been performed with the powerful package LiE \cite{lie}.

%%%%%%%%%%%%%%%%%%%%%%%%%%%%%%%%%%%%%%%%%%%%%%%
\begin{appendix}
\section{Generating function for \texorpdfstring{$SU(3)$}{SU(3)}} \label{append1}
In this section, we derive an analytic solution to eq.~(\ref{exa11}). By invariance of the
Haar measure, $d\mu(g)=d\mu(g^{-1})$, we expect $Z_{SU(3)}$ to be a symmetric function of
$u$ and $v$. Thus, our objective is to solve (\ref{exa11}) and (\ref{exa13})
with the symmetry $Z_{SU(3)}(u,v)=Z_{SU(3)}(v,u)$ and initial condition $Z_{SU(3)}(0,0)=1$. From~\refs{einf23}, we know that
\eqnn{
Z_{SU(3)}(0,v)=\sum_{n=0}^\infty \frac{2}{n!(n+1)!(n+2)!}\,v^{3n}.}
Multiplying~(\ref{exa11}) by $u$ and~(\ref{exa13}) by $v$, the difference of the resulting equations reads
\eqnl{
\big(u^2\partial_u^2 - v^2\partial_v^2 + 4\left(u\partial_u-v\partial_v\right) -
3(u^2\partial_v - v^2\partial_u)\big)Z_{SU(3)} = 0.}{a3sol3}
In principle, it is possible to find a recursive solution analogously to
\refs{exb16} and \refs{exg25} starting from a solution on the locus where the families of characteristics of the differential equation coincide (here on the $u$- and the $v$-axis),
where the differential system turns out to be parabolic. However, in the case of $SU(3)$
we are able to give a closed expression of the solution which allows for an easier
computation of the moments~\refs{einf4}:

Center symmetry~\refs{exa15} as well as the symmetry of $Z_{SU(3)}$ in its two arguments
suggests to introduce new coordinates $x=uv$ and $y=u^3+v^3$. As a function of these
variables, $Z_{SU(3)}$ has to satisfy
\eqnl{(3y\partial_y^2 + 6\partial_y + 2x\partial_x\partial_y-\partial_x)Z_{SU(3)} = 0,}{a3sol5}
which can easily be solved by a power series expansion,
\eqnl{Z_{SU(3)}(x,y) = \sum_{m,n=0}^\infty a_{mn}x^m y^n}{a3sol7}
with recursion relation
\eqnl{(3n+3+2m)n a_{mn} = (m+1)a_{m+1,n-1}.}{a3sol9}
The solution to \refs{a3sol7} and \refs{a3sol9} with the condition
\eqnl{Z_{SU(3)}(x=0,y) = \sum_{n=0}^\infty \frac{2}{n!(n+1)!(n+2)!)}\,y^n\equiv\sum_{n=0}^\infty a_{0n}y^n}{a3sol11}
is given by
\eqnl{Z_{SU(3)}(u,v) = \sum_{m,n=0}^\infty\frac{2}{n!(m+n+1)!(m+n+2)!}{3(m+n+1)\choose m}(uv)^m \big(u^3+v^3\big)^n.}{a3sol13}
This solution also satisfies the original differential equations~(\ref{exa11}) and (\ref{exa13}). It encodes information about how many invariants there are in the tensor product $3^{\otimes p}\otimes\bar{3}^{\otimes q}$ of $SU(3)$ representations. General results for 
integrals over $SU(3)$ matrix elements can be found in~\cite{eriksson}.
%%%%%%%%%%%%%%%%%%%%%%%%%%%%%%%%%%%%%
%%%%%%%%%%%%%%%%%%%
\section{Tables of moments for the rank \texorpdfstring{$2$}{2} groups}\label{append2}
With the recursion relations \refs{moma1}, \refs{moma3}, \refs{momb1}, \refs{momb3}, \refs{momg1}, and \refs{momg3}, we determined the moments
\eqnl{
t_{mn}=\int_G \dmur (g) \chi^m_{[1,0]}(g)\chi^n_{[0,1]}(g),}{momap1}
for the groups with rank $2$ for small $m$ and $n$.
Here $\chi_{[1,0]}$ and $\chi_{[0,1]}$ are the characters of the fundamental
representations with highest weights $\mu_1\equiv [1,0]$ and $\mu_2\equiv [0,1]$.

\textbf{For SU(3)} the lowest moments of $\chi_3^m\chi_{\bar 3}^n$ are
\eqnn{
\begin{array}{c|rrrrrrrrrrr}
m\backslash n&0 &1 &2&3 &4 &5 &6 &7 &8 &9 &10\\ \hline
0& 1 &0 & 0& 1 & 0 & 0 & 5 & 0 & 0 & 42 & 0\\
1& 0 &1 & 0& 0 & 3 & 0 & 0 &21  &0  & 0 & 210\\
2& 0 & 0& 2& 0 & 0 & 11 & 0 & 0 & 98 &0  & 0\\
3& 1 &0 &0  &6 &0 &0  &47 &0 &0  &498&0\\
4&0 &3 &0&0 &23 &0 &0 &225 &0 &0 &2709\\
5& 0&0 & 11 &0 &0 & 103 &0  & 0 & 1173 &0 &0\\
6& 5&0 &0&  47 &0 &0 &513 &0 &0 &6529&0\\
7& 0 & 21  &0&0 & 225 &0 &0 &2761 &0 &0 &38265\\
8& 0&0 & 98&0 &0 & 1173 &0 &0 & 15767 &0 &0\\
9& 42& 0 & 0& 498 &0 &0 & 6529 &0 &0 & 94359 &0\\
10& 0& 210 &0&0 &2709 &0 &0 &38265 &0 &0 &586590\\
\end{array}
}
\textbf{For Spin(5)} the lowest moments of $\chi_5^m\chi_{4}^n$ are
\eqnn{
\begin{array}{c|rrrrrrrrrrr}
m\backslash n&0 &1 &2&3 &4 &5 &6 &7 &8 &9 &10\\ \hline
0&   1 &0 & 1& 0 & 3 & 0 & 14 & 0 & 84 & 0 & 594\\
1&   0 &0 & 1& 0 & 5 & 0 & 30 &0  &210  & 0 & 1650\\
2& 1 & 0& 2& 0 & 11 & 0 & 75 & 0 & 580 &0  & 4917\\
3& 0 &0 &4 &0 &27 &0  &205 &0 &1714  &0&15435\\
4&3 &0 &10&0 &73 &0 &600 &0 &5338 &0 &50506\\
5& 1&0 & 26 &0 &211 & 0 &1852  & 0 & 17342 &0 &171022\\
6& 15&0 &75&  0 &645 &0 &5970 &0 &58350 &0&596085\\
7& 15 & 0  & 225&0 & 2061 &0 &19950 &0 &202230 &0 &2129719\\
8& 105&0 & 715&0 &6837 & 0 &68730 &0 & 718928 &0 &7774600\\
9& 190& 0 & 2347& 0 &23403 &0 & 243050 &0 &2612796 & 0 &28922112\\
10& 945& 0 &7990&0 &82301 &0 &879204&0 &9681144 &0 &109404729\\
\end{array}
}
Both tables nicely display the constraints from center symmetry.\\
\textbf{For G$_2$} the lowest moments of $\chi_7^m\chi_{14}^n$ are
\eqnn{
\begin{array}{c|rrrrrrrrr}
m\backslash n&0 &1 &2&3 &4 &5 &6 &7 &8 \\ \hline
0&   1 &0 & 1& 1 & 5 & 16 & 80 & 436 & 2786 \\
1&   0 &0 & 0& 1 & 6 & 40 & 260 &1785  &12852  \\
2& 1 & 1& 3& 10 & 45 & 236 & 1421 & 9444 & 67852 \\
3& 1 &2 &7 &32 &170 &1016  &6637 &46656 &348553  \\
4& 4 &9 &33&151 &817 &4984 &33357 &240181 &1835171\\
5& 10 & 30 &126 &641 & 3728 &23986  & 167080 &1241285& 9727650\\
6& 35&120 &545&  2932 &17827 & 118945&854135 &6511050 & 52159514\\
7& 120 & 476  & 2359&13517 & 86171 &596686&4415055 & 34500369& 282217558\\
8& 455&2002 & 10626& 64078&425194& 3041241 &23115050 &184754906 & 1540766892 \\
\end{array}
}
\end{appendix}

%%%%%%%%%%%%%%%%%%%%%%%%%%%%%%%%%%%%%%%%%%%%%%%%%%%%%%%%%%%%%%%%%%%%%%%%%%%%%%%%%%

%%%%%%%%%%%%%%%%%%%%%%%%%%%%%%%%%%%%%%%%%%%%%%%%%%%%%%%%%%%%%%%%%%%%%%%%%%%%%%%%%%%%%%%%%%%%%%%%%%

\end{document}